\documentclass[]{revtex4-2}

\usepackage{graphicx}

\begin{document}

\title{Resistivity in Quantum Vortex Liquid of Clean Two-Dimensional Superconductor}

\author{Naratip Nunchot and Ryusuke Ikeda}
\affiliation{Department of Physics, Kyoto University, Kyoto 606-8502, Japan} 

\begin{abstract}
Motivated by a recent controversy on a possible quantum phase in thin films of relatively clean superconductors under an out-of-plane magnetic field, the quantum fluctuation effects on the phase diagram and the resistivity are reexamined. It is argued that most of features seen in the corresponding resistivity data in relatively clean systems reported recently are explained within the present theory, and that the fan-shaped resistivity curves, suggestive of the presence of a superconductor to insulator transition at zero temperature, in the vortex liquid regime is a consequence of the insulating behavior of the Aslamasov-Larkin fluctuation conductivity in the quantum regime.
\end{abstract}

\maketitle

\section{Introduction}
In thin films of type II superconductors under a magnetic field perpendicular to the plane, the resistivity often shows a behavior insensitive to the temperature $T$ over wide field and temperature ranges \cite{Kapi1,Kapi2}. Possibilities of a novel two-dimensional (2D) quantum phase based on this quantum metallic behavior have been discussed repeatedly over the past two decades \cite{Kapi3,nature,Nojima1}. However, it has been clarified recently that most of the $T$-independent behavior of the resistivity is removed by adequately filtering external radiation from the film sample \cite{Tamir,India}, strongly suggesting that external noise has created the quantum metallic behavior in experiments. The presence of a quantum metal state has been still argued in some recent experimental works on relatively clean systems, i.e., with {\it weak} disorder \cite{Nojima1,Ienaga,Masonjyanaihou,Shahar23}. Since a nearly flat resistivity curve is seen even in the temperature range of the same order as the mean field $T_c$ in some film samples, such a peculiar resistive behavior cannot be due to the randomness or 
the sample disorder which becomes more effective at lower temperatures. In addition, a crossing behavior leading to assuming the presence of a superconductor to insulator quantum transition (SIT) at zero temperature \cite{MPAF,HP} is seen at relatively higher fields in samples of relatively clean films \cite{Ienaga,Masonjyanaihou}. Then, one might wonder what the flat resistivity curve appearing in clean samples in lower fields than the {\it apparent} SIT field implies. 

In the present work, the quantum superconducting (SC) fluctuation effects on the resisitivity in clean and 2D superconductors are reexamined by performing a detailed analysis within the framework of the renormalized fluctuation theory \cite{IOT,IOT2}. It was argued in a previous theoretical work of one of the present authors \cite{RI96b} that, based on a dimensional analysis, the melting curve $H_m$ of the 2D vortex lattice becomes insensitive to $T$ at low enough temperatures due to the quantum SC fluctuation, and that, in such a quantum regime, the vortex flow resistance in a narrow field range close to $H_m$ is also insensitive to $T$ and takes a value of the order of the quantum resistance $R_q=\pi e^2/2 \hbar = 6.45$(k$\Omega$). However, this explanation on the crossing behavior seen in the field dependence of the resistivity curves seems to be inconsistent with the observation of the {\it apparent} SIT behavior in a couple of experiments \cite{Ienaga,Masonjyanaihou} where the crossing of the resistivity is seen in a much higher field than the nominal vortex lattice melting field at low temperatures. Below, the vortex lattice melting transition line will be first examined without resorting to the rough argument \cite{RI96b} and by comparing the free energy of the renormalized fluctuation of the SC order parameter with that of the vortex lattice corrected by the Gaussian fluctuation \cite{Hikami}. In contrast to the previous estimate of the quantum melting line \cite{RI96b}, the resulting melting field $H_m$ grows upon cooling everywhere at nonzero temperatures, while $H_m(T=0)$ can take a much lower value than $H_{c2}(T=0)$, and the resulting quantum vortex liquid regime becomes well-defined \cite{Blatter,RI96b}. Next, the in-plane resistivity computed within the renormalized fluctuation theory is examined in a consistent way with the calculation of the melting line. Bearing in our mind that the characteristic features of the resistivity curves in the quantum regime seem to depend on the details of the materials, the resistivity curves will be discussed by focusing on the two extremely different cases: One is the case with a moderate strength of the thermal fluctuation and an extremely strong quantum fluctuation, and the other is the case with strong thermal fluctuation and weak quantum fluctuation. In both cases, the crossing behavior of the resistivity leading to erroneously assuming the presence of an SIT at zero temperature appears in a finite temperature range, as a consequence of the fact that the Aslamasov-Larkin (AL) term of the dc fluctuation conductivity vanishes in the vortex liquid in zero temperature limit \cite{RI96b,RI96a}. The resistivity curve insensitive to $T$ tends to appear more frequently when the thermal fluctuation is stronger. 

This paper is organized as follows. We explain the theoretical treatment used in the present work in sec.2. The resulting numerical results on the phase diagram and the resistivity curves are presented in sec.3. Summary of our results is given and relevance to the experimental data are given in sec.4. 

\section{Theoretical Expressions}
In the unit of $k_{\rm B}=\hbar=1$, we start from the partition function 
\begin{equation}
Z= {\rm Tr}\exp(-{\cal S}). 
\end{equation}
Here, in the high field approximation where the pair field $\psi({\bf r})$ consists 
only of the lowest Landau level (LLL) modes $\psi_0({\bf r})$, the action ${\cal S}$ expressing the Ginzburg-Landau (GL) model takes the form \cite{RI96a,RI96b,IOT2} 
\begin{eqnarray}
{\cal S} &=& \sum_{\omega, p} (s \omega^2 + \gamma_0|\omega| + \varepsilon_0) \, |{\tilde \psi}_0(p; \omega)|^2 \nonumber \\
&+& \frac{g}{2 d \beta^2} \int_0^{\beta} d\tau \int d^2r \, |\psi_0({\bf r}, 
\tau)|^4. 
\label{GLaction}
\end{eqnarray}
Here, the order parameter field was rescaled so that the dependences on the film thickness $d$ and the temperature $T=\beta^{-1}$ appear only in the quartic term. Further, the order parameter field was expanded in terms of the normalized eigen functions $u_p({\bf r})$ in LLL in the manner $\psi_0({\bf r}, \tau) = \sum_{p, \omega} {\tilde \psi}_0(p, \omega) e^{-i \omega \tau} u_p({\bf r})$, $\omega$ is the Matsubara frequency for bosons, and $p$ measures the macroscopic degeneracy in LLL. The microscopic $T$ and $H$ dependences of the positive coefficients $s$, $\gamma_0$, and $g$ are, for simplicity, neglected, and the bare mass $\varepsilon_0$ will be assumed to be linearly dependent on $H$ and $T$ like 
\begin{equation}
\varepsilon_0 = t-1+h, 
\end{equation}
where $h=H/H_{c2}(0)$, and $t=T/T_{c0}$. The mean field $H_{c2}(T)$ line is given by $\varepsilon_0=0$. Further, since the $\omega^2$ term in the  action ${\cal S}$ was introduced only to cut off an inessential divergence in the frequency summation, the coefficient $s$ is assumed to be small so 
that $s \ll \gamma_0^2$. 

The simplest approximation describing reasonably the fluctuation renormalization is the Hartree approximation which is reached through the self-consistent replacement 
\begin{equation}
|\psi_0|^4 \,\,\,\, \to \,\,\,\, 4 \, \langle |\psi_0|^2 \, \rangle |\psi_0|^2 
\end{equation}
in the quartic term, where $\langle \,\,\,\,\, \rangle$ denotes the statistical average within the Hartree approximation. Then, the fluctuation propagator ${\cal G}_0(p, \omega)=\langle |{\tilde \psi}_0(p, \omega)|^2 \rangle$ is given by $1/[r_0 + \gamma_0|\omega| + s \omega^2]$, where 
\begin{equation}
r_0 = \varepsilon_0 + \frac{g h}{\pi \xi_0^2 d} \, \beta^{-1} \sum_\omega \frac{1}{r_0 + \gamma_0|\omega| + s \omega^2},  
\end{equation}
where $\xi_0$ is the coherence length in zero temperature limit. 
Note that, according to the BCS theory \cite{deGennes}, the mode-coupling strength $g$ is a positive constant of the order of $(N(0) T_{c0}^2)^{-1}$, where $N(0)$ is the density of states of the quasiparticles on the Fermi energy in the normal state. To rewrite the frequency summation into a tractable form, the spectral representation \cite{RI96b,Tsuneto}
\begin{equation}
\frac{1}{r_0 + \gamma_0|\omega| + s \omega^2} = \frac{1}{\pi} \int_{-\infty}^{\infty} du \, \frac{\rho(r_0; u)}{u - {\rm i}\gamma_0 \omega} 
\end{equation}
will be used, where 
\begin{equation}
\rho(r; u) = \frac{u}{(u^2 + (a r)^2)((us/\gamma_0^2)^2 + a^{-2})}. 
\label{spect} 
\end{equation}
This expression (7) of the spectral function is valid when $r < \gamma_0^2/(4s)$. Then, the coefficient $a$ in eq.(\ref{spect}) is given by $a^{-1}= (1 + \sqrt{1 - 4sr/\gamma_0^2})/2$. Since we are interested in the region below $H_{c2}(T)$-line where $r_0 \ll 1$, the coefficient $a$ will be replaced by unity in the ensuing expressions.  
Therefore, we will use hereafter the following self-consistent relation on the renormalized mass $r_0$ of the LLL fluctuation 
\begin{equation}
r_0 = \varepsilon_0 + \frac{2 \varepsilon_{\rm G}^{(2)} \, h}{\pi \gamma_0 T_{c0}} \int_0^\infty du \, {\rm coth}\biggl(\frac{u}{2 \gamma_0 T} \biggr) \, \frac{u}{(u^2 + r_0^2)( 1 + (s u/\gamma_0^2)^2)}, 
\label{renmass}
\end{equation}
where $H_{c2}(0)$ is the depairing field in zero temperature limit, 
\begin{equation}
\varepsilon_{\rm G}^{(2)} = \frac{g T_{c0}}{2 \pi \xi_0^2 d} 
\end{equation}
is the Ginzburg-number in 2D, and the identity 
\begin{equation}
{\rm coth}\biggl(\frac{u}{2 \gamma_0 T} \biggr) = 2 \gamma_0 T \sum_\omega \frac{1}{u - {\rm i}\gamma_0 \omega} 
\end{equation}
was used. Note that eq.(\ref{renmass}) can be regarded as being a definition of $\varepsilon_0(r_0)$ as a function of $r_0$. Then, we have 
\begin{equation}
\frac{\partial \varepsilon_0(r_0)}{\partial {r_0}} = 1 + \frac{2 \varepsilon_{\rm G}^{(2)} \, h}{\pi \gamma_0 T_{c0}} \int_0^\infty du \, {\rm coth}\biggl(\frac{u}{2 \gamma_0 T} \biggr) \frac{2 r_0 u}{(u^2 + r_0^2)^2}.
\label{derr}
\end{equation}  

\subsection{Free energy}
Next, the expressions on the free energy density will be derived. Using the identity on the fluctuation free energy $F_>$ 
\begin{equation}
\frac{\partial F_>}{\partial \varepsilon_0} = \sum_{p, \omega} {\cal G}_0(p, \omega), 
\end{equation} 
the fluctuation free energy density $f_>$ in the vortex liquid regime of a SC thin film with thickness $d$ is simply given by 
\begin{equation}
f_> = \frac{h}{2 \pi^2 \xi_0^2 d \gamma_0} \int_{r_c}^{r_0} d\mu \int_0^\infty dx \, {\rm coth}\biggl(\frac{x}{2 \gamma_0 T} \biggr) \, \rho_\mu(x) \frac{\partial \varepsilon_0(\mu)}{\partial \mu}, 
\end{equation}
where the prefactor proportional to $h$ arises from the degeneracy in LLL. Then, $f_>$ will be expressed in terms of eq.(\ref{derr}) as 
\begin{equation}
f_>=f_{\rm G}(r) + f_{\rm H}, 
\label{fabove}
\end{equation}
where 
\begin{equation}
f_{\rm H} = \frac{h}{2 \pi^2 \xi_0^2 d \gamma_0} \int_{r_c}^{r_0} d\mu \int_0^\infty dx \, {\rm coth}\biggl(\frac{x}{2 \gamma_0 T} \biggr) \, \rho_\mu(x) \biggl(\frac{\partial \varepsilon_0(\mu)}{\partial \mu} - 1 \biggr). 
\end{equation}
The cut-off $r_c$ will be determined in examining $f_{\rm G}(r_0)$ 
(see below). 

Regarding the remaining term $f_{\rm G}(r_0) = f_> - f_{\rm H}$ which is nonvanishing even when $g=0$, i.e., even in the absence of the mode-couplings, the $\mu$-integral will be performed firstly. Then, $f_{\rm G}(r_0)$ 
takes the form 
\begin{equation}
f_{\rm G}(r_0) = \frac{H}{\phi_0 \pi d \gamma_0} \int_0^\infty dx \frac{1}{1+(sx/\gamma_0^2)^2} {\rm coth}\biggl(\frac{x}{2 \gamma_0 T} \biggr) \biggl[ {\rm tan}^{-1}\biggl(\frac{x}{r_c} \biggr) - {\rm tan}^{-1}\biggl(\frac{x}{r_0} \biggr) 
\biggr]. 
\end{equation}
Here, to determine the cut-off $r_c$, we take the thermal limit of eq.(\ref{renmass}) in which ${\rm coth}(x/(2 \gamma_0 T))$ is replaced by $2 \gamma_0 T/x$. By comparing it with the corresponding result in ref.16, $h T \, {\rm ln}(r_0/r_c)/(2 \pi \xi_0^2 d)$, the cut-off will be chosen hereafter as 
\begin{equation} 
r_c = \pi \gamma_0 T. 
\end{equation}

On the other hand, by making use of eq.(\ref{renmass}) determining the $T$ and $H$ dependences of the renormalized mass $r_0$, $f_{\rm H}$ may be rewritten in the following simpler form
\begin{equation}
f_{\rm H} = - \frac{1}{4g} (r - \varepsilon_0 \,)^2. 
\end{equation}

The free energy derived above can be used as the SC fluctuation contribution to the free energy in the normal phase. To determine the 2D quantum melting transition line, the corresponding free energy density $f_<$ in the vortex lattice phase corrected by the Gaussian fluctuations is needed. Within the GL approach, the contribution of the shear elastic energy is smaller \cite{RI90} in the order of the magnitude than that of the amplitude (or, Higgs) mode and hence, will be simply neglected. Then, $f_<$ becomes 
\begin{equation} 
f_< = - \frac{1}{2 \beta_{\rm A}} \biggl( \frac{\varepsilon_0^2}{g}  - \frac{1}{\sqrt{2}} f_{\rm G}(-2 \varepsilon_0) \biggr), 
\label{fbelow}
\end{equation}
where $\beta_{\rm A}$ is the Abrikosov factor $1.1596$ of the triangular lattice. Using these expressions, the transition line of the 2D vortex lattice melting occurring through not only the thermal but also the quantum fluctuations of the SC order parameter is determined by the relation $f_>=f_<$. 

\subsection{Fluctuation Conductivity}
The fluctuation conductivity in the moderately clean case is dominated by the Aslamasov-Larkin (AL) term of the conductivity due to the renormalized SC fluctuation which is expressed in dc limit by \cite{RI96b}
\begin{equation}
d R_q \sigma_{\rm AL} = 2 T r_1^2 \sum_\omega \biggl[ {\cal G}_0(\omega) {\cal G}_1(\omega) ( \, \gamma_0 {\cal G}_0(\omega) + \gamma_1 {\cal G}_1(\omega) \, ) - \frac{\gamma_0^2 [{\cal G}_0(\omega)]^2 + \gamma_1^2 [{\cal G}_1(\omega)]^2}{\gamma_0 r_1 + \gamma_1 r_0} \biggr], 
\label{cond}
\end{equation}
where ${\cal G}_n(\omega)= 1/(\gamma_n|\omega| + r_n)$ (see also eq.(15) of Ref.14 and eq.(21) of Ref.22). The time scale $\gamma_1$ is the counterpart in the second lowest ($n=1$) Landau level (LL) fluctuation of $\gamma_0$ of the LLL fluctuation, and the tiny $\omega^2$ term introduced in eq.(5) as a cut-off term for the frequency summation is unnecessary in obtaining $\sigma_{\rm AL}$ and hence, has been neglected in eq.(20). As shown previously \cite{IOT2}, the renormalized mass $r_1$ of the $n=1$ LL fluctuation is renormalized to be $2h$ deep in the vortex liquid regime in the Hartree approximation. Hereafter, the relations $r_1=2h$ and $\gamma_0=\gamma_1$ will be assumed for simplicity in our numerical analysis. 

\section{Numerical Results}

Now, we will explain typical examples of the resistivity curves following from eqs.(\ref{renmass}) and (\ref{cond}) together with the corresponding phase diagrams which follow from eqs.(\ref{fabove}) and (\ref{fbelow}). Below, the coefficient of the $\omega^2$ term of eq.(\ref{GLaction}) which plays the role of a cutoff on the dissipative dynamics will be chosen as $s = (10^{-6} \gamma_0)^2$ throughout this paper. 

In our work, the DOS and Maki-Thompson fluctuation terms of the conductivity are not taken into account from the outset based on the well-known fact \cite{Varlamov,Galitski,Nunchot} that, in clean limit, those terms and the subleading contribution of the Aslamasov-Larkin term cancel with one another in 2D systems with no Pauli paramagnetic depairing. For this reason, the total dimensionless conductivity $R_q d \sigma_{\rm tot}$ is assumed hereafter to be given by the sum of the leading contribution of the Aslamasov-Larkin term in dc limit, eq.(\ref{cond}), and the dimensionless normal conductivity $d R_q \sigma_{\rm N}$. Regarding $\sigma_{\rm N}$, the same model as used in Ref.14, $d R_q \sigma_{\rm N} = (1 + (8 \pi)^{-1} {\rm ln}(T_{c0}/T))^{-1}$, will be used here to describe a weakly insulating resistivity curve in the normal state of a couple of materials \cite{Ienaga,Masonjyanaihou}. 

\begin{figure}

\includegraphics{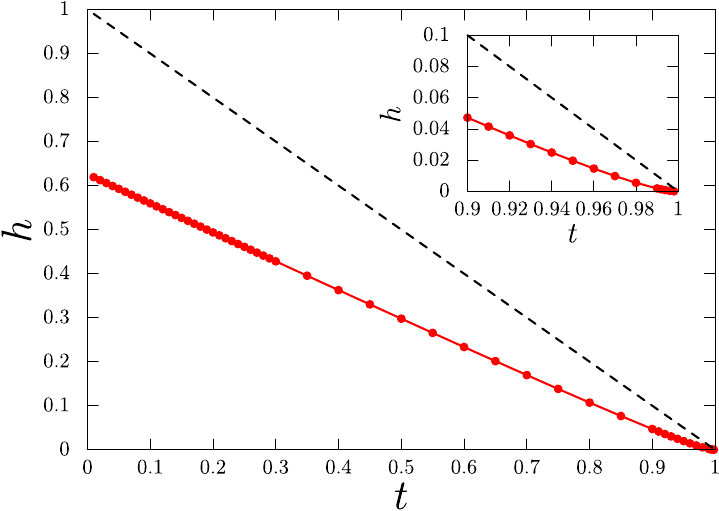}
\caption{(Color online)  Field ($h=H/H_{c2}(0)$) v.s. temperature ($t=T/T_{c0}$) phase diagram of a clean 2D superconductor with moderately strong thermal fluctuation ($\varepsilon_{\rm G}^{(2)}=2.0 \times 10^{-4}$) and unusually strong quantum fluctuation ($\hbar/(\gamma_0 T_{c0})= 10^2$). The red symbols and line express the vortex lattice melting curve, and the black dashed line is the $H_{c2}(T)$-line. The Inset is presented to clarify the details of the melting curve close to $T_{c0}$. }
\label{fig.1}
\end{figure}

\begin{figure}

\includegraphics{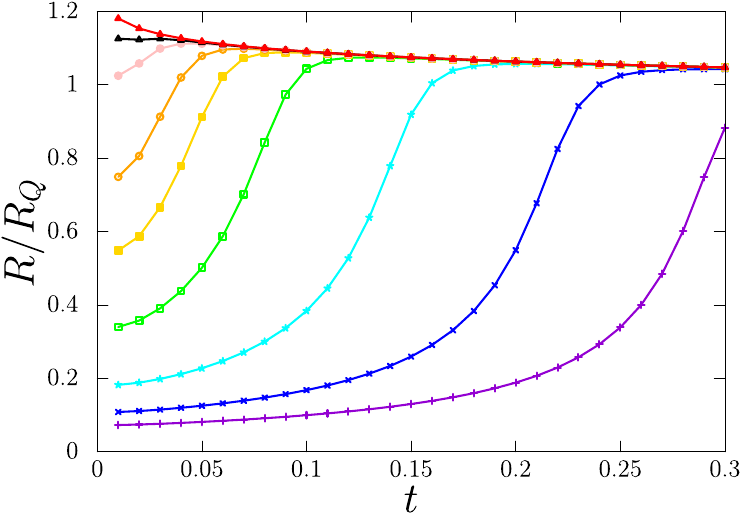}
\caption{(Color online)  Temperature dependences of the dimensionless resistance $R/R_q$ at various magnetic fields $h=0.5$, $0.55$, $0.6$, $0.65$, $0.66$, $0.67$, $0.68$, $0.69$, and $0.7$ from the bottom to the top in the situation of Fig.1. When $h \leq 0.55$, the resistance curve is flattened below the melting curve, while it does not become flat in higher fields where the system is in the vortex liquid regime everywhere. Due to the strong quantum fluctuation, the temperature at which the resistance drops lies far below the $H_{c2}(T)$-line, and a fan-shaped $T$-dependence of the resistance curves is visible around $h=0.69$. }
\label{fig.2}
\end{figure}

To clarify what are typical consequences originating from strong quantum SC fluctuations, typical results following from two highly different sets of the parameter values will be compared with each other. Below, the strengths of the thermal fluctuation and the quantum fluctuation will be measured, respectively, by $\varepsilon_{\rm G}^{(2)} = [\lambda(0)]^2/(d \Lambda(T_{c0}))$ and $\hbar/(\gamma_0 k_{\rm B}T_{c0})$, where $\lambda(0)$ is the magnetic penetration depth at $T=0$, $\Lambda(T)=\phi_0^2/(16 \pi^2 T)$, and $\phi_0$ is the flux quantum \cite{deGennes}. Here, we have used the relation between $g$ and $\lambda(0)$ in the BCS theory \cite{deGennes}. 

First, the results of the phase diagram in a case with moderately strong thermal fluctuation and unusually strong quantum fluctuation are shown in Fig.1 where $\hbar/(\gamma_0 k_{\rm B}T_{c0})= 100$ and $\varepsilon_{\rm G}^{(2)}=2.0 \times 10^{-4}$. This $\varepsilon_{\rm G}^{(2)}$-value corresponds to, e.g., the set of the parameter values $T_{c0}=10$(K), $d=25$(A), and $\lambda(0)=330$(A). It is found that the melting field $H_m(T)$ is linear in the temperature over a wide field range except close to $T_{c0}$. Close to $T_{c0}$, the quantum fluctuation is negligible so that $H_m(T)$ in the present LLL-GL approach obeys the 2D LLL scaling \cite{com} $H_m(T) \simeq (T_{c0} - T)^2$ (see the Inset of Fig.1). Such a large deviation of the melting line from its LLL scaling behavior over the wide field range is a consequence of the strong quantum fluctuation in this case, and the $T=0$ melting field $H_m(0)$ becomes $0.62 H_{c2}(0)$. 

Figure 2 expresses the resistivity curves $\rho(T)$ at various magnetic fields, $H/H_{c2}(0) = 0.5$, $0.55$, $0.6$, $0.65$, $0.66$, $0.67$, $0.68$, $0.69$, and $0.7$. The two curves in lower fields than $H_m(0)$ are found to become flat, i.e., insensitive to $T$, below the melting line, while each of other curves in $H > H_m(0)$ simply shows a drop at a temperature without a clear flat portion accompanied. We note that each temperature $T_d$ at which the resistivity starts to drop is much lower than $T_{c2}(H)$ corresponding to the mean field $H_{c2}(T)$-line. For instance, at $H=0.66 H_{c2}(0)$, $T_d/T_{c0}=0.06$, while $T_{c2}/T_{c0}=0.34$ \cite{comhc2}. Such a large deviation of $T_d$ from $T_{c2}$ is a consequence of strong reduction of $\sigma_{\rm AL}$ (eq.(20)) due to the unusually strong quantum fluctuation assumed in Figs.1 and 2. On the other hand, the flat (i.e., metallic) portion is not clearly seen in those resistivity curves. As will be stressed below, it appears that the flat portion does not become remarkable as far as the {\it thermal} fluctuation is not strong enough. Nevertheless, as a consequence of the strong quantum superconducting fluctuation, the so-called fan-shaped $T$-dependence of the resistivity curves which often leads to assuming the presence of a superconductor to insulator transition (SIT) at $T=0$ is seen in the field range $(H_m(0) <) \, 0.67 H_{c2}(0) < H < 0.7 H_{c2}(0)$ in spite of the absence of a quantum continuous transition. It seems that these resistivity curves are qualitatively similar to the data in Refs. 5, 8, and 9. Of course, it should be noted that those resistive behaviors explained above in $H > 0.6 H_{c2}(0)$ are not their genuine low $T$ results. Since there are no quantum transitions above $H_m(0)$ in the present clean limit, all curves of the normalized resistance $1/(d R_q \sigma_{\rm tot})$ in $H > H_m(0)$ start to grow at much lower temperatures than $0.01 T_{c0}$ and reduce to their normal values $1/(d R_q \sigma_{\rm N})$ on approaching $T=0$ reflecting the vanishing of $\sigma_{\rm AL}$ at $T=0$ \cite{RI96a,RI96b}. 

\begin{figure}
\includegraphics{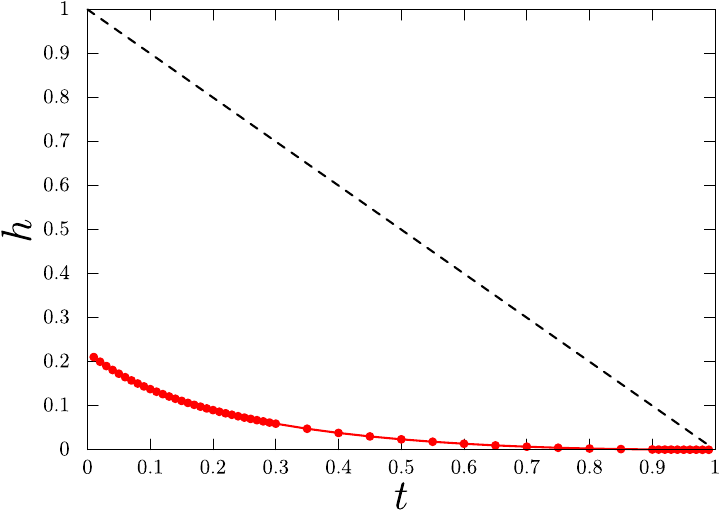}
\caption{(Color online)  The $h$ v.s. $t$ phase diagram, corresponding to Fig.1, obtained in the case with quite a strong thermal fluctuation strength ($\varepsilon_{\rm G}^{(2)}=0.12$) and a quantum fluctuation strength with a moderate magnitude $\hbar/(\gamma_0 T_{c0})= 1.0$.}
\label{fig.3}
\end{figure}

\begin{figure}
\includegraphics{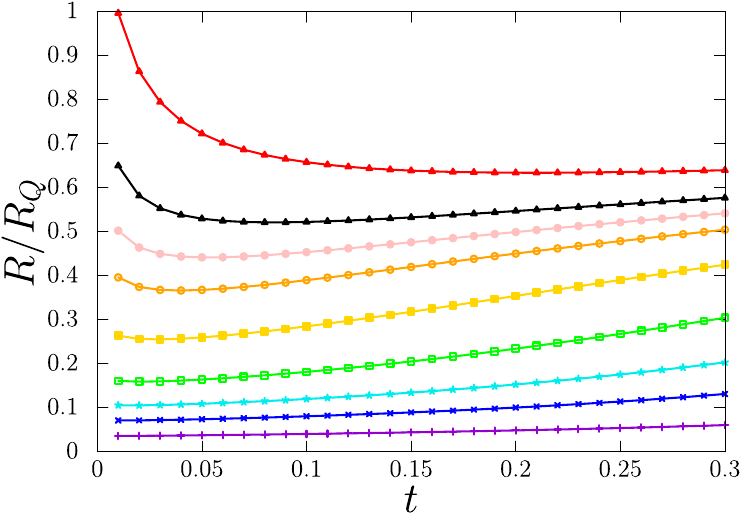}
\includegraphics{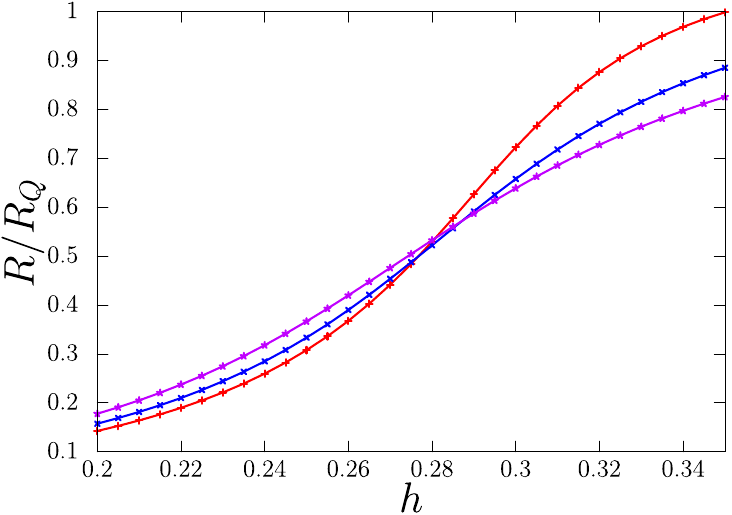}
\caption{(Color online) (Upper) Temperature dependences of the dimensionless resistance $R/R_q$ at various magnetic fields $h=0.1$, $0.15$, $0.18$, $0.21$, $0.24$, $0.26$, $0.27$, $0.28$, and $0.3$ from the bottom to the top in the situation of Fig.3. Note that the resistivity curve below $H_m(T)$ becomes flat at lower temperatures. (Lower)  Crossing behavior seen among the resistance curves of the upper figure in the temperature region $0.04 < t < 0.16$. }
\label{fig.4}
\end{figure}

Next, the case with exceptionally strong thermal fluctuation and a moderate strength of the quantum fluctuation will be considered. In Fig.3 and 4, we have used $\varepsilon_{\rm G}^{(2)} = 0.12$, corresponding to, e.g., the set of the parameter values $T_{c0}=30$(K), $d=5$(A), and $\lambda(0)=2000$(A), and $\hbar/(\gamma_0 k_{\rm B}T_{c0})= 1.0$. As Fig.3 shows, the vortex liquid regime is expanded particularly at higher temperatures reflecting the large $\varepsilon_{\rm G}^{(2)}$, and the melting curve is bent upwardly at low enough temperatures reflecting the relatively weaker quantum fluctuation. Nevertheless, the $H_m(0)/H_{c2}(0)$-value is remarkably low, and, in $H > 0.2 H_{c2}(0)$, we have only the vortex liquid regime at any temperature. 

In Fig.4, the corresponding resistivity curves are shown. It is noticeable that nearly flat resistivity curves are seen over a wide field range. This is a consequence of the strong {\it thermal} fluctuation assumed here. In particular, the flat resistivity curves appear in fields below $H_m(0)$, i.e., the fluctuating vortex solid phase. Here, we stress that, in 2D case, the freezing from the vortex liquid to the vortex solid tends not to be reflected in the resistivity curve. It has been recently clarified through a detailed diagrammatic analysis \cite{Nunchot2} that this feature on the resistivity in 2D case has a theoretical foundation. 

Even in the resistivity data of Fig.4, a crossing behavior of the resistivity curves is seen at nonzero temperatures. As is seen in the lower figure of Fig.4, the resistivity curves in the field range $0.2 H_{c2}(0) < H < 0.34 H_{c2}(0)$ obey an approximate crossing behavior around $H=0.28 H_{c2}(0)$ in the temperature range $0.04 T_{c0} < T < 0.16 T_{c0}$. Again, this crossing behavior never implies the presence of a genuine quantum transition 
and is merely a reflection of the insulating behavior of the fluctuation conductivity \cite{RI96a,RI96b} arising from the vanishing of eq.(\ref{cond}) 
at $T=0$. 

\section{Summary and Discussion} 

In this work, we have examined possible field v.s. temperature phase diagrams and the corresponding resisitivity curves to be seen in thin films of clean superconductors under a magnetic field perpendicular to the two-dimensional plane. Since a moderately strong fluctuation has been assumed in obtaining those figures, the field range of our interest in which the vortex lattice melting occurs at zero temperature is low enough to neglect the paramagnetic pair-breaking effect. In this situation, the fluctuation conductivity in clean 2D superconductors is given by only the well-known Aslamasov-Larkin term \cite{Varlamov,Galitski,Nunchot}. For this reason, we have been able to assume that, even at low temperatures, the total conductivity is the sum of a quasiparticle contribution and the conventional fluctuation conductivity following from a time-dependent GL dynamics. We note that, in a moderately dirty system \cite{IOT2} case, the sum of the Maki-Thompson and DOS terms of the fluctuation conductivity has a contribution leading to a negative magnetoresistance in the fluctuation regime \cite{Galitski,Gant}. Therefore, the absence of such a negative magnetoresistance would play a key role in judging whether the present theory is applicable to experimental data of the resistivity or not. 

The resistivity curves obtained based on the renormalized fluctuation theory \cite{IOT,IOT2} are highly dependent on the relative magnitude of the quantum fluctuation to the thermal one. When the thermal fluctuation is of a moderate strength, enhanced quantum fluctuation tends to create fan-shaped resistivity curves $R(T)$, often leading to erroneously assuming the presence of a quantum SIT, in the quantum vortex liquid but far above the vortex lattice melting field in $T=0$ limit. This type of resistivity data have been reported in several works \cite{Ienaga,Masonjyanaihou}. 
Further, even in quite a different case where the thermal fluctuation is quite strong, while the quantum fluctuation has a moderate strength, the resistive behavior suggestive of the presence of an apparent SIT is visible in the experimentally measurable temperature range. We conclude that, except the observations in dirty systems \cite{MPAF,HP,IOT2,Gant}, the SIT behavior of the resistivity in relatively clean systems is a consequence of the insulating behavior \cite{RI96a,RI96b,Nunchot} of the Aslamasov-Larkin fluctuation conductivity in dc limit in the quantum regime. 
 
In the present work, any pinning effect arising from some randomness or defects in the SC material has been neglected. In analyzing resistivity data in thin films, the resistivity drop upon cooling at intermediate temperatures is often modelled according to the empirical thermal activation (TA) (or the so-called Arrhenius) formula. Within the GL model, this TA behavior may be conveniently incorporated as an exponential growth in the inverse temperature $T^{-1}$ of the coefficient $\gamma_1$. As far as the vortex lattice melting transition does not occur due to weak disorder in the material, the present fluctuation theory can be used even for the lower temperature region, in which a flat resistive behavior may be seen, than the region of the vortex liquid in which the TA behavior is seen. In fact, it is interesting to regard a (if any) flat resistivity curve as a consequence of a competition between the insulating fluctuation conductivity \cite{RI96a,RI96b} and an increase of $\gamma_1$ on cooling. 

\begin{acknowledgments}
The present work was supported by a Grant-in-Aid for Scientific Research [Grant No.21K03468] from the Japan Society for the Promotion of Science. 


\end{acknowledgments}




\begin{thebibliography}{9}
\bibitem{Kapi1} D. Ephron, A. Yazdani, A. Kapitulnik, and M. R. Beasley, Phys. Rev. Lett. {\bf 76}, 1529 (1996). 
\bibitem{Kapi2} N. Mason and A. Kapitulnik, Phys. Rev. Lett. {\bf 82}, 5341 (1999). 
\bibitem{Kapi3} J. A. Chervenak and J. M. Valles, Jr., Phys. Rev. B {\bf 61}, 9245(R) (2000). 
\bibitem{nature} Y. Qin, C. L. Vicente, and J. Yoon, Phys. Rev. B {\bf 73}, 100505(R) (2006). 
\bibitem{Nojima1} Y. Saito, T. Nojima, and Y. Iwasa, Nature Comm. {\bf 9}, 778 (2018). 
\bibitem{Tamir} I. Tamir, A. Benyamini, E. J. Telford, F. Gorniaczyk, A. Doron, T. Levinson, D. Wang, F. Gay, B. Sacepe, J. Hone, K. Watanabe, T. Taniguchi, C. R. Dean, A. N. Pasupathy, and D. Shahar, Sci. Adv. {\bf 5}, 3826 (2019).  
\bibitem{India} Surajit Dutta, Indranil Roy, Soumyajit Mandal, John Jesudasan, Vivas Bagwe, and Pratap Raychaudhuri, Phys. Rev. B {\bf 100}, 214518 (2019). 
\bibitem{Ienaga} K. Ienaga, T. Hayashi, Y. Tamoto, S. Kaneko, and S. Okuma, Phys. Rev. Lett. {\bf 125}, 257001 (2020). 
\bibitem{Masonjyanaihou} Wei Liu. LiDong Pan, Jiajia Wen, M. Kim, G. Sambandamurthy, and N. P. Armitage, Phys. Rev. Lett. {\bf 111}, 067003 (2013). 
\bibitem{Shahar23} A. Haug and D. Shahar, arXiv: 2305.1593. 
\bibitem{MPAF} M. P. A. Fisher, Phys. Rev. Lett. {\bf 65}, 923 (1990). 
\bibitem{HP} A. F. Hebard and M. A. Paalanen, Phys. Rev. Lett. {\bf 65}, 927 (1990). 
\bibitem{IOT} R. Ikeda, T. Ohmi, and T. Tsuneto, J. Phys. Soc. Jpn. {\bf 58}, 3770 (1989). 
\bibitem{IOT2} R. Ikeda, J. Phys. Soc. Jpn. {\bf 72}, 2930 (2003). 
\bibitem{RI96b} R. Ikeda, Int. J. Mod. Phys. B {\bf 10}, 601 (1996). 
\bibitem{Hikami} S. Hikami, A. Fujita, and A. I. Larkin, Phys. Rev. B {\bf 44}, 10400(R) (1991). 
\bibitem{Blatter} G. Blatter, B. Ivlev, Y. Kagan, M. Theunissen, Y. Volokitin, and P. Kes, Phys. Rev. B {\bf 50}, 13013 (1994). 
\bibitem{RI96a} R. Ikeda, J. Phys. Soc. Jpn. {\bf 65}, 33 (1996). 
\bibitem{deGennes} P. G. de Gennes, {\it Superconductivity of Metals and Alloys} (Addison Wesley, 1989). 
\bibitem{Tsuneto} E. Abrahams and T. Tsuneto, Phys. Rev. B {\bf 11}, 4498 
(1975). 
\bibitem{RI90} G. Eilenberger, Phys. Rev. {\bf 164},628 (1967). 
\bibitem{Nunchot} N. Nunchot, D. Nakashima, and R. Ikeda, Phys. Rev. B {\bf 105}, 174510 (2022). 
\bibitem{Varlamov} D. V. Livanov, G. Savona, and A. A. Varlamov, Phys. Rev. B {\bf 62}, 8675 (2000).  
\bibitem{Galitski} V. M. Galitski and A. I. Larkin, Phys. Rev. B {\bf 63}, 174506 (2001). 
\bibitem{com} For simplicity, effects of the higher LL modes making the vertical portion of the melting curve in lower fields in the field v.s. temperature phase diagram will be neglected. See T. Saiki and R. Ikeda, Phys. Rev. B {\bf 83}, 174501 (2011). 
\bibitem{comhc2} In Ref.5, the $H_{c2}(T)$-curve has been determined based on the LLL scaling relation \cite{IOT} formulated by {\it neglecting} the quantum fluctuation in spite of the fact that the resistivity curves show the fan-shaped SIT behavior. In the phase diagram proposed in Ref.5 (Fig.4 there), the correct $H_{c2}(T)$-curve {\it must} lie at a much {\it higher} temperature at least in higher fields, and it seems to us that their erroneous determination of the $H_{c2}(T)$-curve has led to their argument on the presence of a quantum Griffiths state which should not appear in cleaner systems of a type studied in Ref.5. 
\bibitem{Nunchot2} N. Nunchot and R. Ikeda, unpublished. 
\bibitem{Gant} V. F. Gantmakher, M. V. Golubkov, V. T. Dolgopolov, G. E. Tsydynzhapov, and A. A. Shashkin, JETP Letters {\bf 68}, 344 (1998). 


\end{thebibliography}
\end{document}